\documentclass[preprint,prc,showpacs,preprintnumbers,
               superscriptaddress,amsmath,amssymb,floatfix]{revtex4}

\usepackage{graphicx,epsfig,latexsym,amssymb}
\usepackage{multirow,amsmath,array,booktabs}
\usepackage{dcolumn}
\usepackage[section]{placeins}

\renewcommand{\emph}[1]{\textit{\textbf{#1}}}

\begin{document}

\title{Theoretical study of the two-proton halo candidate $^{17}$Ne including
contributions from resonant continuum and pairing correlations}

\author{Shi-Sheng Zhang}

\email{zss76@buaa.edu.cn}

\affiliation{School of Physics and Nuclear Energy
Engineering, Beihang University, Beijing 100191, China}
\affiliation{Institute of Theoretical Physics, Chinese Academy of
              Sciences, Beijing 100190, China}

\author{En-Guang Zhao}
\email{egzhao@mail.itp.ac.cn}
\affiliation{Institute of Theoretical Physics, Chinese Academy of
             Sciences, Beijing 100190, China}
\affiliation{Center of Theoretical Nuclear Physics, National Laboratory
              of Heavy Ion Accelerator, Lanzhou 730000, China}

\author{Shan-Gui Zhou}
\email{sgzhou@itp.ac.cn}
\affiliation{Institute of Theoretical Physics, Chinese Academy of
              Sciences, Beijing 100190, China}
\affiliation{Center of Theoretical Nuclear Physics, National Laboratory
              of Heavy Ion Accelerator, Lanzhou 730000, China}
\date{\today}

\begin{abstract}

With the relativistic Coulomb wave function boundary condition,
the energies, widths and wave functions of the single proton resonant orbitals
for $^{17}$Ne are studied by the analytical continuation of the coupling constant (ACCC) approach
within the framework of the relativistic mean field (RMF) theory.
Pairing correlations and contributions from the single-particle resonant orbitals in the continuum
are taken into consideration by the resonant Bardeen-Cooper-Schrieffer (BCS) approach,
in which constant pairing strength is used. It can be seen that the fully self-consistent
calculations with NL3 and NLSH effective interactions
mostly agree with the latest experimental measurements,
such as binding energies, matter radii, charge radii and densities.
The energy of $\pi$2s$_{1/2}$ orbital is slightly higher than that of $\pi$1d$_{5/2}$ orbital, and
the occupation probability of the $(\pi$2s$_{1/2})^2$ orbital is about $20\%$,
which are in accordance with the shell model calculation and three-body model estimation. 
\end{abstract}


\maketitle

\section{Introduction}
The study of unbound or loosely bound exotic nuclei casts a new
light on the discovery of nuclear halo structure~\cite{TA85}. The
valence nucleons can be readily scattered into single-particle
resonant orbitals in the continuum. Due to the small binding energy
and small or no centrifugal barrier conditions, the valence nucleons
tunnel out of the potential barrier to long distances with an
extended density tail to form the so-called halos. Therefore, the
resonant orbitals in the continuum and the coupling between bound
states and the continuum near the threshold play an important role
in the description of halo phenomena~\cite{Bulgac80,Doba84}.
In most calculations, the continuum is replaced by a set of positive energy orbitals
without the contributions of the widths, determined by solving
the Hartree-Fock-Bogoliubov (HFB) or
relativistic-Hartree-Bogoliubov (RHB) equations in coordinate space
and with box boundary conditions~\cite{Doba96,MR96}.
Taking into account of the contributions from bound states and resonant orbitals in the continuum,
the resonant Bardeen-Cooper-Schrieffer approach based on the relativistic mean field theory (RMF-rBCS)~\cite{San00,San01,San03,Cao04,Zhang09}
is applied to describe the properties of neutron-rich nuclei.

Recently, neutron halos have been prominently observed in experiments~\cite{TA85,Aumann00,Ye05,Mueller07,Smith08}
and widely explored theoretically~\cite{Renzz95,ME98,MR98,ZhangHQ01,ZhouSG10}.
However, studies on proton halos are still scarce because of the existence of the Coulomb barrier~\cite{ZhangHQ04,Ye09,Mayg10},
and even more so for two-proton halos~\cite{Renzz96}.
As one of the promising two-proton halo candidates,
$^{17}$Ne recently attracts experimental~\cite{Kanungo03,Geithner08,Tanaka2010-prc} and
theoretical~\cite{Sagawa11,Sagawa10,Fortune01,Zhukov05,Zhanghy03,Ma11} attention.
The most precise mass and charge radii measurements of
$^{17-22}$Ne have been performed with Penning trap mass spectrometry and collinear
laser spectroscopy~\cite{Geithner08}. Furthermore,
the reaction cross sections for $^{17}$Ne are measured by the transmission method
and the corresponding density distribution is deduced from
a modified Glauber-type calculation~\cite{Tanaka2010-prc}.

The improved experimental measurements for $^{17}$Ne
make it possible to better compare with theoretical calculations.
Using a Woods-Saxon potential together with Coulomb interaction in three-body model,
the ground state properties of $^{17}$Ne are analyzed~\cite{Sagawa10}
in which the pairing correlation parameters are adjusted to
reproduce the threshold of two-proton emission.
The root mean square (rms) radius for $^{17}$Ne is estimated to be 2.62 fm and the s-orbit occupation probability is about $15\% $.
It is also useful to make similar calculations for $^{17}$Ne in our fully self-consistent model,
including proton resonant orbitals contributions, pairing correlations and Pauli block.

The paper is organized as follows. Numerical details, results and discussion are given in Section II.
The conclusions and outlook for future work are outlined in Section III.

\section{Numerical details, results and discussion}

The single-particle (s.p.) energies E$_B$ and wave functions for bound states are extracted
from the RMF approach. With the relativistic Coulomb wave functions boundary condition~\cite{Zhang07},
the s.p. energies E$_R$, widths $\Gamma$, and wave functions for proton resonant orbitals
are extracted from the analytical continuation of the coupling constant (ACCC)~\cite{kukulin89} approach
within the framework of the RMF theory. Using E$_B$, E$_R$ and $\Gamma$, we solve the BCS gap equations
to obtain the Fermi energy, pairing gap, the occupation probabilities of
bound and resonant orbitals~\cite{Zhang09}.
With contributions from the relevant s.p. orbitals, the nuclear densities are recalculated
solving the coupled Dirac equations, as well as equations of meson field and photon field.
The total binding energy is recovered from the summation of s.p. energies, meson energies, pairing energies and center-mass corrections.
The next iteration starts with new E$_B$, E$_R$, $\Gamma$ and wave functions for s.p. orbitals extracted from the new potential.
Therefore, the RMF+ACCC+BCS calculations are numerically self-consistent by
an iterated way until convergency. The binding energies
are precise to $10^{-3}$ MeV and the densities to
$10^{-4}$ fm$^{-3}$. Details for the neutron case can be found in Ref.~\cite{Zhang09}.


The pairing window is opened from all the bound states below the Fermi surface
to one harmonic oscillator shell above the Fermi surface in our model,
including the bound proton states 1s$_{1/2}$, 1p$_{3/2}$, 1p$_{1/2}$,
and single proton resonant states 1d$_{5/2}$, 2s$_{1/2}$. The
constant pairing strength $G = C/A$ in the BCS approximation, is
chosen by fitting the odd-even mass difference extracted from three-point formula for neutron pair,
\begin{equation}
\Delta_n = \dfrac{1}{2}[B(Z,N-1)- 2B(Z,N) + B(Z,N+1)],
\label{Constant pairing gaps}
\end{equation}
in which binding energies are taken from Ref.~\cite{Audi03} except for $^{17}$Ne by latest mass measurements~\cite{Geithner08}.
For $^{17}$Ne, $C_n=25.5$ (26.0) can reproduce $\Delta_n = 1.8$ MeV for NL3 (NLSH) effective interaction.
By fitting the odd-even mass difference $\Delta_p = 0.9$ MeV extracted from three-point formula for proton pair,
\begin{equation}
\Delta_p = \dfrac{1}{2}[B(Z-1,N)- 2B(Z,N) + B(Z+1,N)],
\label{Constant pairing gaps}
\end{equation}
$C_p=10.5$ (11.0) for NL3 (NLSH) effective interaction.
Highly excited resonant orbitals with large widths, such as 1d$_{3/2}$
and 1f$_{7/2}$, are ignored in our calculations because of the minor
occupation probability. For odd neutrons, the blocking effect is
considered in the pairing correlations~\cite{Bender00}.
Note that RMF+ACCC+BCS approach is different from the conventional BCS approach,
taking only the real resonant orbitals into consideration and kicking
the spurious continuous orbitals off from the continuum~\cite{Cao04,Zhang09}.
Using this approach, the properties of s.p. bound orbitals and resonant orbitals
in stable and unstable Ni, Zr and Sn isotopes have been well described~\cite{Zhang09,Zhang12-EPJA, Zhang12-PRC}.

The theoretical binding energies for $^{17}$Ne and $^{15}$O are calculated by the
RMF, RMF+BCS and RMF+ACCC+BCS approaches with NL3 and NLSH effective interactions~\cite{Lal97, Sharma93}.
As is well known, the results of the RMF+BCS approach (i.e., the conventional BCS approach) change with the box size
because of the inclusion of the spurious orbitals.
As opposed to the RMF+BCS approach, the results of the RMF+ACCC+BCS approach are invariable for the specified pairing window,
which is from all the bound states below the Fermi surface
to one harmonic oscillator shell resonant orbitals above the Fermi surface.
Therefore, only the binding energies obtained from the RMF and
the RMF+ACCC+BCS approaches are shown in Tab.~I.
The available experimental binding energy for
$^{17}$Ne is 112.9276 MeV and that for $^{15}$O is 111.955 MeV~\cite{Audi03}.
An asterisk denotes the result of the RMF+ACCC+BCS calculation without resonant width contribution.
In this case, the resonant orbitals are simplified by the discrete positive orbitals without the widths.
It can be seen from Tab.~I that the results without width effect are larger
than those with widths contributions because of the over evaluation of the pairing correlation,
which is also mentioned in Ref.~\cite{Cao04}.
The binding energies in the RMF+ACCC+BCS calculations are 1$\%$ larger than experimental data.
The two-proton separation energy S$_{2p}$ for $^{17}$Ne is 0.93 MeV, and the corresponding
theoretical result is 0.2 (0.14) MeV for the NL3 (NLSH) effective interaction.
The values of S$_{2p}$ in our calculations are smaller than data, but those at least tell us
$^{17}$Ne is a weakly bound nucleus, which can not be seen from the RMF calculations.
The binding energy for $^{15}$O is larger than that for $^{17}$Ne
in the RMF calculations, which means $^{17}$Ne is unstable against proton emission
if pairing correlations are not taken into account.
In the RMF+ACCC+BCS calculations, the binding energy for $^{15}$O is less than that for $^{17}$Ne,
which means the valence protons are weakly bound in $^{17}$Ne because of the attractive pairing correlations.
As a consequence, pairing correlations are crucial in the existence of the nuclide
and need to be taken into consideration. The contributions from the resonant orbitals in the continuum
to the pairing are as important as the bound states near the Fermi surface. The width of the resonant orbital
has the effect of weakening the pairing gap and decreasing the binding energy.
All these factors are self-consistently included in the RMF+ACCC+BCS calculations so that
the reasonable description of the binding energy can be given.

To show the contributions from the s.p. energies to the nuclear binding energies,
the single proton orbitals of $^{17}$Ne are displayed in Tab.~II, which are
calculated by the fully self-consistent RMF+ACCC+BCS approach compared with those by RMF approach,
for the NL3 and NLSH effective interactions, respectively.
Note that the positive energies of the s.p. orbitals given by the RMF approach are improper because of the box boundary condition.
The results of the RMF approach are also shown for comparison with the results of the RMF+ACCC+BCS approach.
It can be seen that the energies of the bound states in the RMF calculations are slightly different from those in the RMF+ACCC+BCS calculations
because the self-consistent potential of the latter is not the same as the former.
For protons, the bound orbitals 1s$_{1/2}$, 1p$_{3/2}$, 1p$_{1/2}$, as well as the
resonant orbitals 1d$_{5/2}$ and 2s$_{1/2}$ with narrow widths,
have a similar structure for the NL3 and NLSH effective interactions.
For clarity, the s.p. energies of proton orbitals in $^{17}$Ne
are plotted in Fig.~1 in potential well and barrier, for NL3 effective interaction.
The proton potential with the Coulomb barrier $V_p$ (red curve)
corresponds to the potential for s partial wave;
$V_p+l(l+1)/r^2$ (blue curve) refers to the potential
with the Coulomb barrier and centrifugal barrier for d partial wave.
It can be clearly seen that the 2s$_{1/2}$ orbit (red line) lies below the Coulomb barrier ($l=0$),
and the 1d$_{5/2}$ orbit (blue line) lies below the Coulomb and centrifugal barrier ($l=2$).
Since s and d orbitals with positive energies lie just above
the threshold (E=0) and below the barrier, the two valence protons have the probabilities to tunnel the barrier,
which are proportional to the widths of the resonant orbitals.

Furthermore, the density distributions for the core
nuclide $^{15}$O and two-proton nuclide $^{17}$Ne in our model are
presented in Fig.~2 in logarithm scale, for the NL3 (solid line) and the NLSH (dashed line) effective interactions, respectively.
The error region for the experimental results is displayed by the gray area~\cite{Tanaka2010-prc}.
The radii of $^{15}$O and $^{17}$Ne are also shown in Fig.~2.
It can been seen that the densities from the RMF+ACCC+BCS approach
agree well with the experimental fit results.
The available experimental rms radius of $^{17}$Ne is 2.75(7) fm and experimental charge
radius is 3.042(21) fm. In the RMF+ACCC+BCS calculations,
the rms radius of $^{17}$Ne is
about 2.86 fm for the NL3 effective interaction, and 2.79 fm for the NLSH effective interaction;
the charge radius of $^{17}$Ne is about 3.15 fm for the NL3 effective
interaction, and 3.07 fm for the NLSH effective interaction.
The relative errors are less than 5$\%$.

To investigate detailed contributions to the radii, we plot the
occupation probabilities of the single proton orbitals
1s$_{1/2}$, 1p$_{3/2}$, 1p$_{1/2}$, 1d$_{5/2}$ and 2s$_{1/2}$ in
$^{17}$Ne by the RMF+ACCC+BCS approach in Fig.~3. In our calculations,
the occupation probability of (2s$_{1/2})^2$ is about 20$\%$, and that of (1d$_{5/2})^2$
is about 40$\%$ for the NLSH effective interaction, which is in accordance with the
results in shell-model calculations and three-body model~\cite{Sagawa10,Fortune,Michel}.
In Fig.~4, the real part of the wave functions for the upper component of
proton resonant orbitals 1d$_{5/2}$ and 2s$_{1/2}$ in $^{17}$Ne by the
RMF+ACCC+BCS approach are respectively plotted for the NL3 and NLSH
effective interactions. The extension of the wave functions
for 1d$_{5/2}$ and 2s$_{1/2}$ can be clearly seen at $r>3$ fm, which give the main
contributions to the density of this range for $^{17}$Ne.

Experimental studies~\cite{Warner,Kanungo03,Geithner08,Tanaka2010-prc}
have not been able to confirm the halo structure of $^{17}$Ne,
and the corresponding theoretical studies are still controversial.
The possibility of the halo existence for $^{17}$Ne is discussed by the occupation probabilities
of the resonant orbitals 2s$_{1/2}$ and 1d$_{5/2}$.
With a three-cluster model, Timofeyuk {\it et al.} obtain a (2s$_{1/2})^2$-dominant configuration
~\cite{Timofeyuk}. Some three-body model calculations suggest almost equal occupation probabilities of the
(2s$_{1/2})^2$ and (1d$_{5/2})^2$ orbitals~\cite{Sagawa11,Garrido,Grigorenko}, but some calculations suggest
a (1d$_{5/2})^2$ dominance~\cite{Sagawa10}. From the Coulomb mass
shift, Nakamura {\it et al.} suggest a (2s$_{1/2})^2$ dominance~\cite{Nakamura},
but Fortune {\it et al.} oppositely suggest a (1d$_{5/2})^2$ dominance~\cite{Fortune}.
By calculating the interaction cross section $\sigma_\mathrm I$ with a Hartree-Fock type wave function
and the Glauber model~\cite{Kitagawa}, Kitagawa {\it et al.} also proposed a (1d$_{5/2})^2$ dominance.

In our present scheme, the occupation probability of (1d$_{5/2})^2$ is larger than that of (2s$_{1/2})^2$,
because the $\pi$2s$_{1/2}$ orbital lies slightly above $\pi$1d$_{5/2}$ orbital.
The inclusion of tensor force might alter the present results,
which has been discussed for proton 1d$_{5/2}-$1d$_{3/2}$ splitting decreasing for Ca isotopes in Ref.~\cite{Jiang05}
within the framework of the RMF model, also appearing in Ref.~\cite{Wang11} within the framework of HFB+SLy5+tensor force model.
Moreover, the consideration of pairing correlations is important since the filling near the Fermi surface
is embedded in the continuum so that the density of the energy levels becomes large. In this sense,
the state-dependent pairing gap may be necessary. Such an effort on this point is in process.

\section{Summary}

The energies, widths and wave functions of the single proton resonant orbitals for $^{17}$Ne
are studied by the RMF-ACCC approach with the relativistic Coulomb wave functions boundary condition.
Pairing correlations and contributions from the s.p. resonant orbitals in the continuum
are taken into consideration by the resonant BCS approach,
in which constant pairing strength are used.
The fully self-consistently microscopic RMF+ACCC+BCS calculations
reproduce the main experimental properties of two-proton halo candidate $^{17}$Ne,
such as matter radii, charge radii, densities with the NL3 and
NLSH effective interactions. The occupation probability of the proton resonant orbit
$(\pi$2s$_{1/2})^2$ is in accordance with the results of the shell model and three-body model.
The resonant orbitals in the continuum play an important role in the pairing
correlations as well as those bound states near Fermi surface. Our model considers pairing correlations for
proton-rich nuclei by the coupling between a few narrow resonant orbitals and bound orbitals, which can be
safely extrapolated to describe exotic nuclei. Since many nuclei are deformed,
it is desirable to include the deformation effect to study the properties of the deformed nuclei.
Preliminary attempts have been made in a deformed Woods-Saxon potential for coupled Dirac equations,
which will be explained in detail independently.

\begin{acknowledgments}

This work was supported partially by the
Fundamental Research Funds for the Central Universities,
Key Laboratory of Micro-nano Measurement-Manipulation and Physics (Ministry of Education),
the National Natural Science Foundation of China under Grant (No.
10875157, 10979066, 11235002, 11035007 and 10605004),
Major State Basic Research Development Program of China
(Grant No. 2007CB815000), Knowledge Innovation
Project of Chinese Academy of Sciences (Grant
Nos. KJCX2-EW-N01 and KJCX2-YW-N32).
The authors express their gratitude to Prof. M. Fukuda for providing the data in Fig. 2.
Helpful discussions with Prof. H. Sagawa, Prof. G. Colo, Prof. Zhongyu Ma, Prof. Yanlin Ye,
Prof. I. Tanihata and Dr. Zhipan Li are
gratefully acknowledged. The author
express her sincere thanks to Dr. D. W. Bardayan for his careful reading for the manuscript.

\end{acknowledgments}

\newpage
\begin{table}
\vspace{0.5cm} \centering\caption{Binding energies of $^{17}$Ne calculated by the RMF and
the RMF+ACCC+BCS (this work) approach with NL3 and NLSH effective
interactions, respectively, and compared with experimental binding
energy ($B(^{17}$Ne)=112.928 MeV, $B(^{15}$O)=111.955 MeV)~\cite{Audi03}.
Two-proton separation energies $S_\mathrm{2p}$ are also shown
for the RMF and the RMF+ACCC+BCS approach.
All energy values are in unit of MeV. Asterisk denotes
the result of the RMF+ACCC+BCS calculations without resonant widths contributions.}
\vspace{0.5cm}
\begin{tabular}{ccccccccc}
\hline
 &  & approach &$ B_{\bf NL3}$&$B_{\bf NL3}-B_{\bf Exp.}$&$B_{\bf NLSH}$&$B_{\bf NLSH}-B_{\bf Exp.}$\\
 \hline
&$^{17}$Ne&RMF+ACCC+BCS &114.50 &1.57  &113.68   &0.75\\
&         &RMF+ACCC+BCS*&114.84 &1.91  &114.01   &1.08\\
&         &RMF       &112.69 &$-$0.24 &111.86   &$-$1.07\\
\midrule \hline
&$^{15}$O&RMF+ACCC+BCS &114.30 &2.345  &113.54   &1.585\\
&         &RMF       &114.29 &2.33 &113.52   &1.56\\
\midrule \hline
&$S_\mathrm{2p}$&RMF+ACCC+BCS&0.20&        &0.14& \\
&         &RMF        &$-$1.60&       &$-$1.66& \\
\hline
 \end{tabular}
\label{tab-1}
\end{table}

\begin{table}[h]
\vspace{0.5cm} \centering\caption{Single proton energy orbitals
for $^{17}$Ne in the RMF and the RMF+ACCC+BCS calculations with NL3 and NLSH effective interactions,
respectively. All energy values are in unit of MeV.} \vspace{0.5cm}
\vspace{0.5cm}
\begin{tabular}
{c@{\hspace{0.5cm}}ccc@{\hspace{0.8cm}}ccc}\toprule\hline
\multirow{2}{1cm}{$\pi nl_j$}&\multicolumn{3}{c}{NL3}&\multicolumn{3}{c}{NLSH}\\
                             &RMF&\multicolumn{2}{c}{RMF+ACCC+BCS}&RMF&\multicolumn{2}{c}{RMF+ACCC+BCS}\\
                             &E&E&$\Gamma$&E&E&$\Gamma$\\
\midrule \hline
2s$_{1/2}$  & 0.81    & 0.73     &0.02   &1.24     &  1.17   &0.18  \\
1d$_{5/2}$  & 0.49    & 0.53     &0.003  &0.61     &  0.65   &0.0026 \\
1p$_{1/2}$  & $-$9.12 &$-$8.94    &0     &$-$9.23  & $-$9.09   &0     \\
1p$_{3/2}$  & $-$15.53&$-$15.36   &0     &$-$15.98 & $-$15.84  &0     \\
1s$_{1/2}$  & $-$35.71&$-$35.43   &0     &$-$36.31 & $-$36.08  &0     \\
\bottomrule \hline
\end{tabular}
\label{tab-2}
\end{table}

\newpage

\begin{figure}[!t]
\centering \vskip -1cm
\includegraphics[scale=0.65] {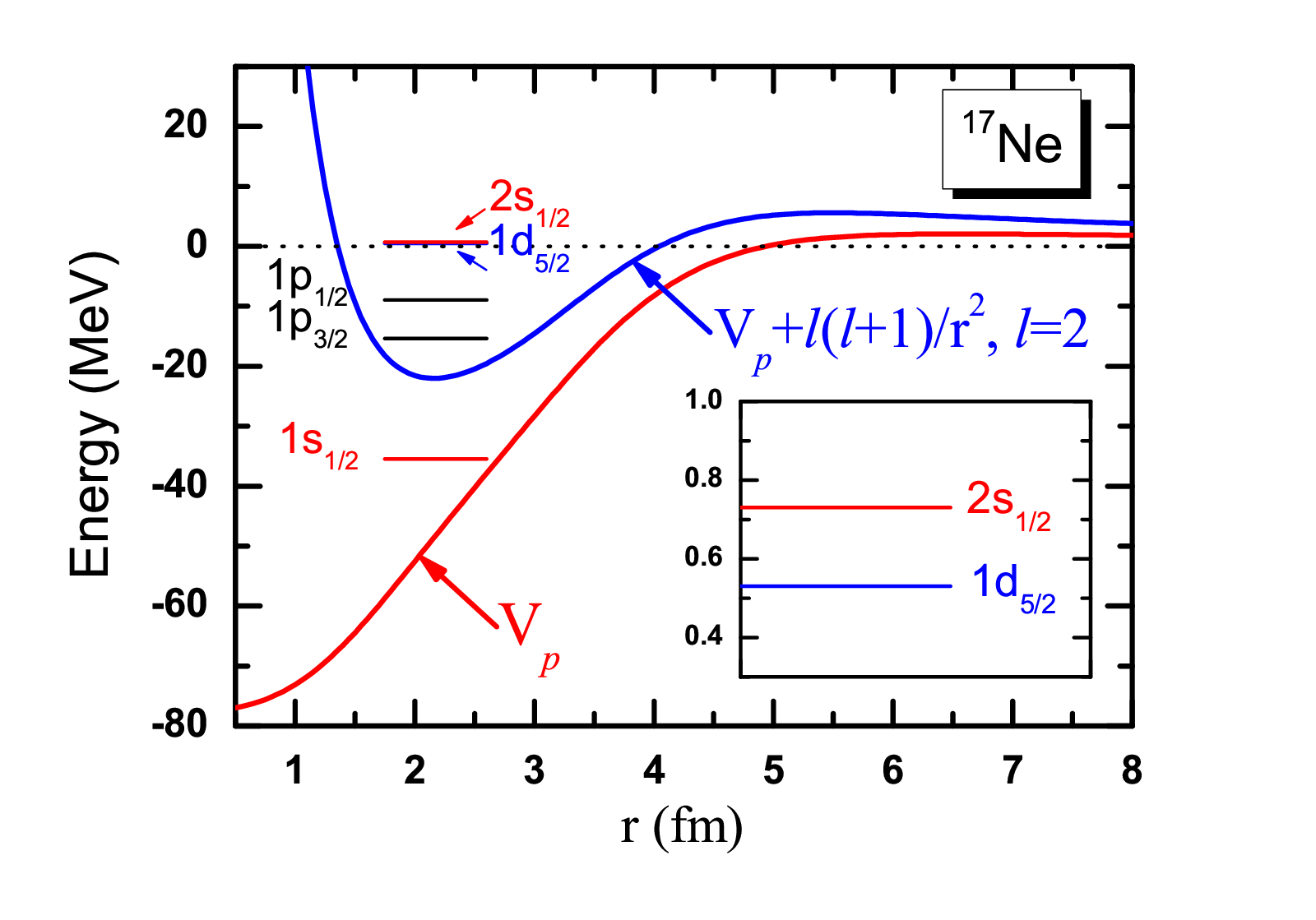}
\caption {(Color online) Single proton orbitals in $^{17}$Ne calculated by the
RMF+ACCC+BCS approach for the NL3 effective interaction.}
\label{Fig1}
\end{figure}

\begin{figure}[!t]
\centering \vskip -1cm
\includegraphics[scale=0.65] {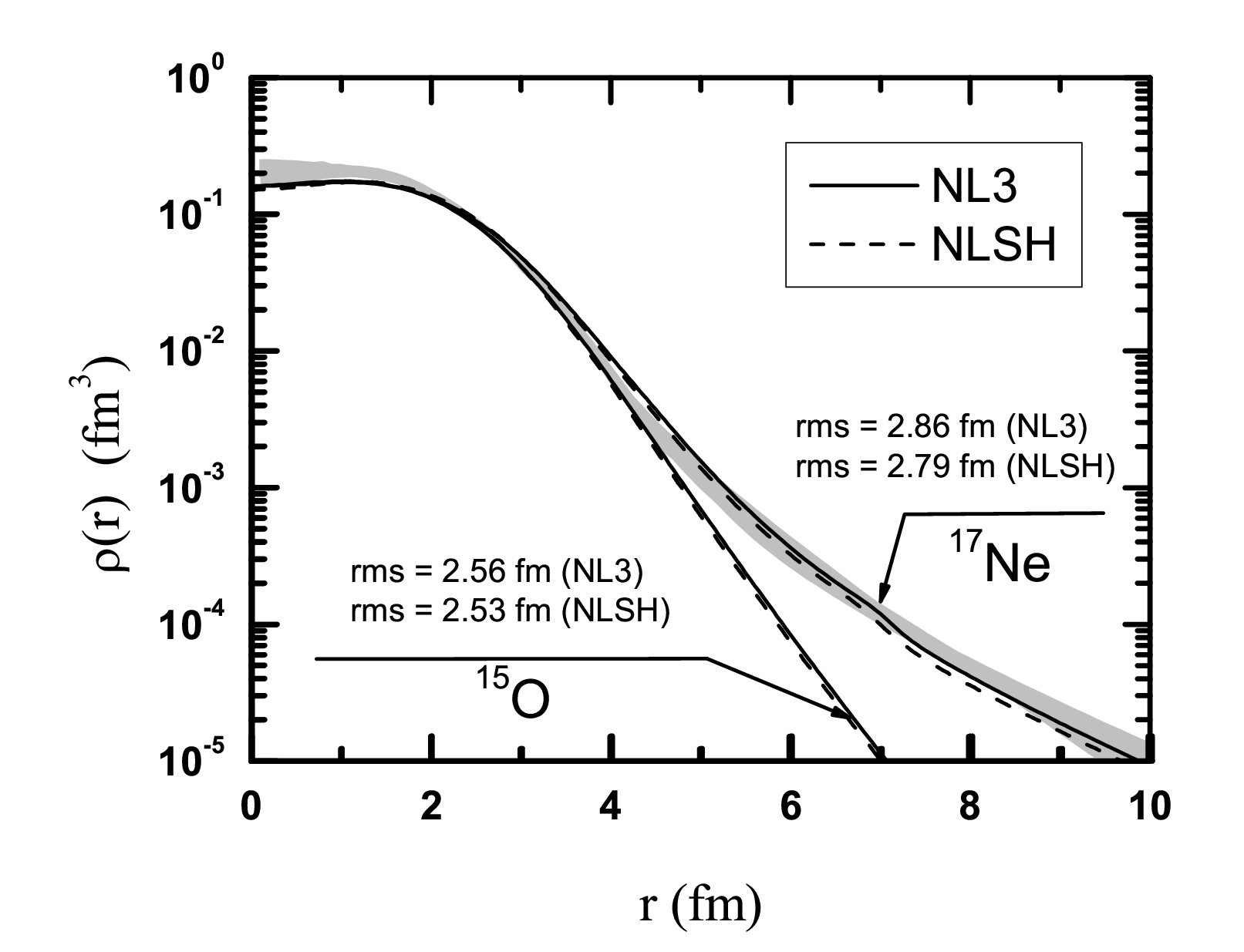}
\caption {Density distributions in logarithm scale for the core nuclide
$^{15}$O and $^{17}$Ne in the RMF+ACCC+BCS calculations with the NL3
(solid line) and the NLSH (dashed line) effective interactions,
respectively. The gray region corresponds to the error region for experimental results~\cite{Tanaka2010-prc}.}
\label{Fig2}
\end{figure}

\begin{figure}[!t]
\centering \vskip -1cm
\includegraphics[scale=0.65] {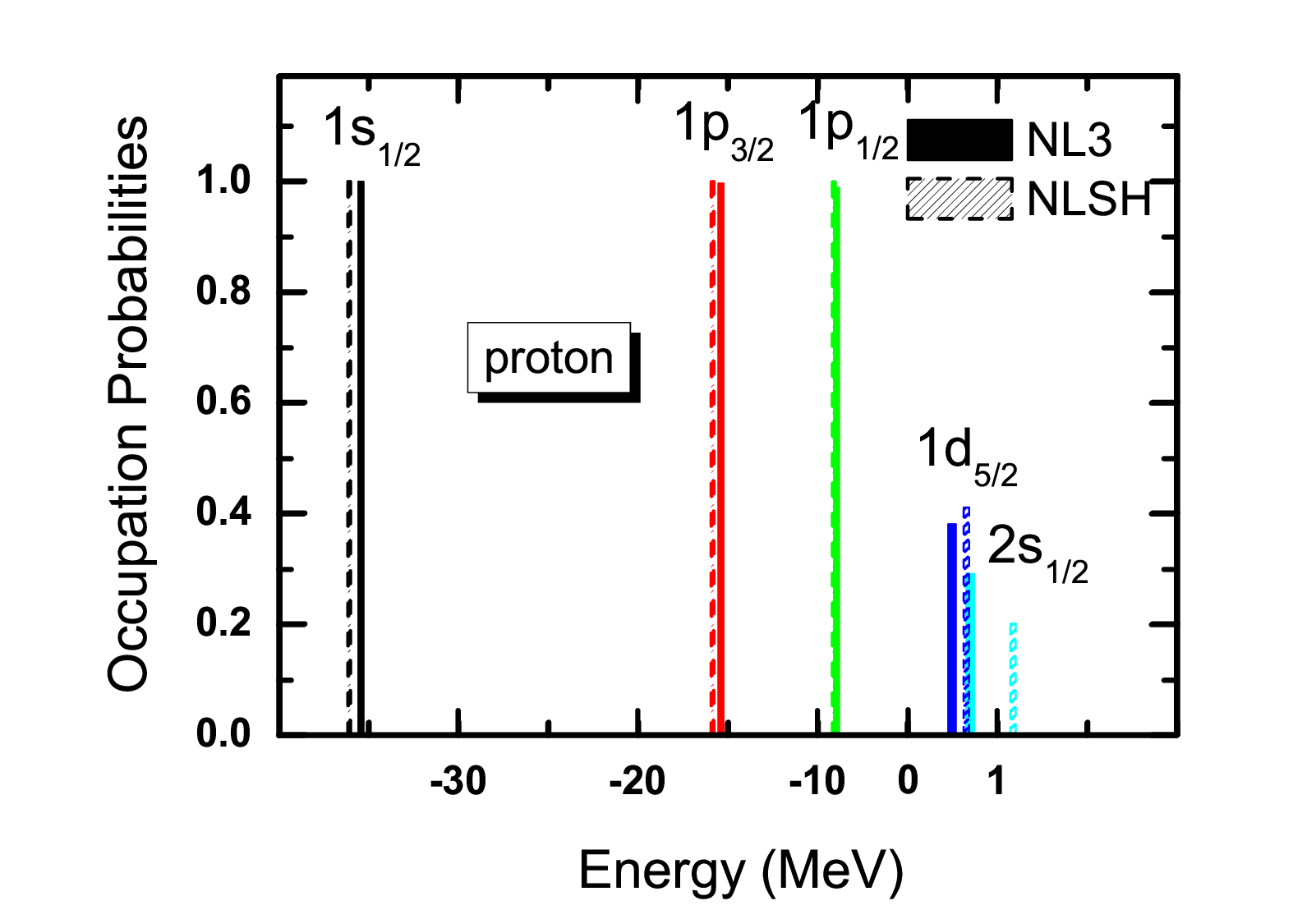}
\caption {(Color online) Occupation probabilities of the proton s.p.
orbitals 1s$_{1/2}$, 1p$_{3/2}$, 1p$_{1/2}$, 1d$_{5/2}$ and 2s$_{1/2}$
in $^{17}$Ne by the RMF+ACCC+BCS approach.}
\label{Fig3}
\end{figure}

\begin{figure}[!t]
\centering \vskip -1cm
\includegraphics[scale=0.65] {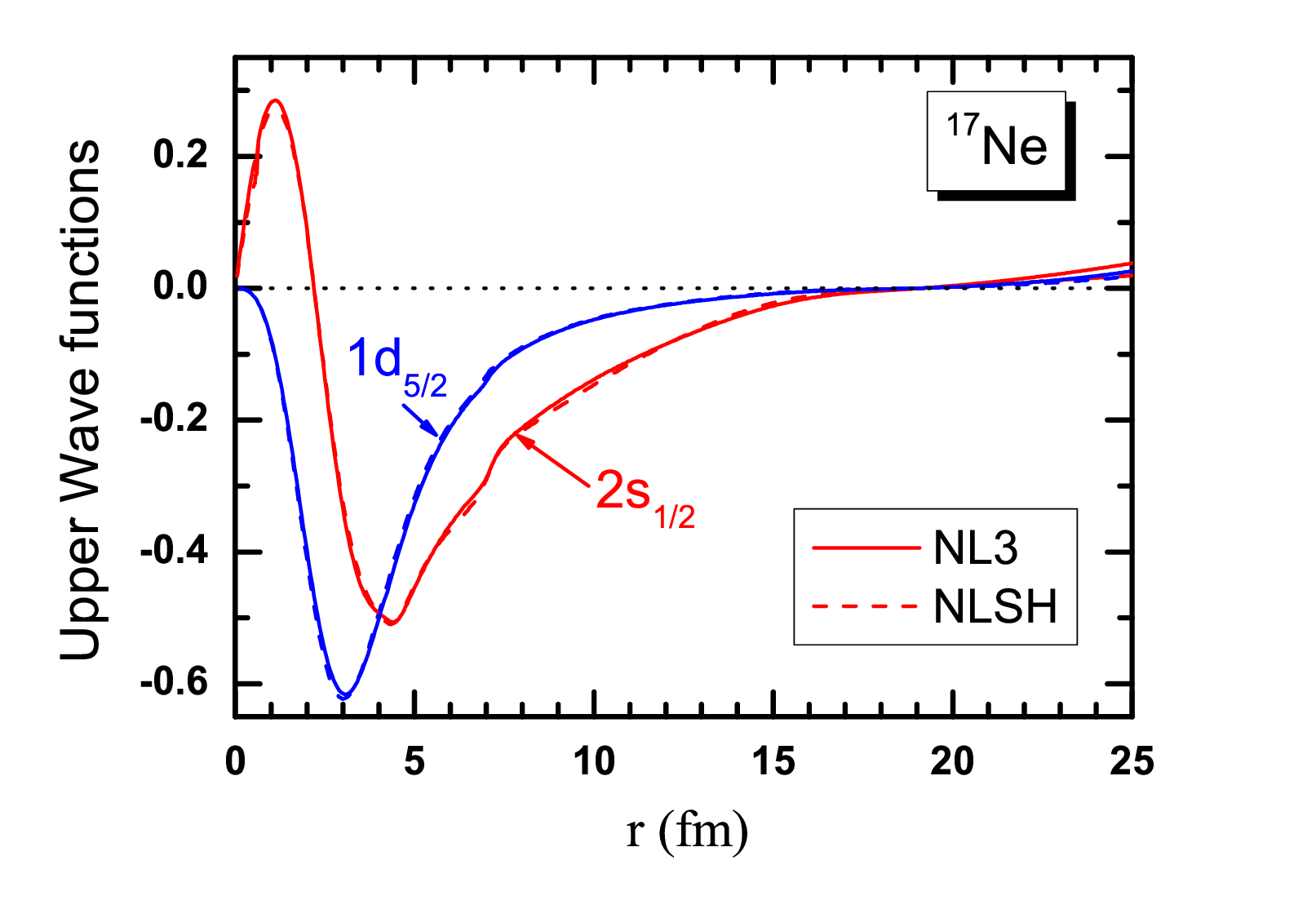}
\caption {(Color online) Upper wave functions for proton resonant orbitals 1d$_{5/2}$
and 2s$_{1/2}$ in $^{17}$Ne by the RMF+ACCC+BCS approach.}
\label{Fig4}
\end{figure}

\end{document}